# Decomposition of Power Flow Used for Optimizing Zonal Configurations of Energy Market

Michał Kłos, Karol Wawrzyniak and Marcin Jakubek
National Centre for Nuclear Research, Świerk Computing Centre, Warsaw, Poland
cst@cis.gov.pl

*Abstract*— Zonal configuration of energy market is often a consequence of political borders. However there are a few methods developed to help with zonal delimitation in respect to some measures. This paper presents the approach aiming at reduction of the loop flow effect – an element of unscheduled flows which introduces a loss of market efficiency. In order to undertake zonal partitioning, a detailed decomposition of power flow is performed. Next, we identify the zone which is a source of the problem and enhance delimitation by dividing it into two zones. The procedure is illustrated by a study of simple case.

## I. INTRODUCTION

The shape of the future bidding zone configuration is a subject of the debate among several institutions determining energy market policy in Europe. One of the main reasons is the urgent need for a close integration of the energy markets of EU members into one structure governed by a common power exchange algorithm – the Market Coupling (MC) [1]. The zonal structure has already been introduced in some European countries; zonal market works for instance in Central-Western Europe (CWE) and among three of CEE countries. These regions are expected to be merged with several independent state markets to form one pan-European market. The shape of current bidding zones (BZ) follows the actual international borderlines. For the reasons discussed further it does not constitute the optimal solution [2]. Therefore, European Network of Transmission System Operators for Electricity (ENTSO-E), which is responsible for introducing zonal model, has recently initiated the process of bidding zones' revision [3,4].

One of the most important problem concerning Transmission System Operators relates to network congestions. In 2011 Agency for the Cooperation of Energy Regulators (ACER) has published Framework Guidelines on Capacity Allocation and Congestion Management for Electricity (CACM). This framework gives strict recommendations that bidding zones should be defined in a way that provides efficient management of congestions. All known model-based approaches to zonal delimitation aim at setting the borders in the manner that congested lines become inter-zonal connections. There are two main groups of methods capable of achieving that goal. The first is based on Locational Marginal Prices [5-9] and aims at aggregation of nodes characterized by similar cost of energy delivered to the node in the nodal model. Second class of algorithms aggregates nodes characterized by similar Power Transfer Distribution Factors in respect to overloaded lines [10-12]. However, the existence of structural congestions is not the only aspect of efficient bidding zone delimitation. CACM states that efficient bidding zone configuration should minimize adverse effects of internal transactions on other BZs. Hence, another important issue that has to be addressed during the process of optimal delimitation is related to unscheduled flows. These flows introduce distortions to neighboring systems decreasing their efficiency and creating potential safety vulnerabilities. The thermal limits of a grid demand reserving sufficient transmission capacity in the form of reliability margin, whereas the magnitude of unscheduled flows can reach even 1000 MW injected into adjacent system without market-based agreements [2, 3, 13]. In some cases this leaves no exchange capability for actual power trade between systems.

The unscheduled flows are defined as the difference between market-driven (scheduled) and actual, physical power flow [13]. There are two components of such flow (i) loop flow, consisting of power transmitted through neighboring zone due to some internal (intra-zonal) transactions and (ii) transit flows which occur when inter-zonal exchange affects third party grids. It is worth mentioning that introducing Flow Based MC changes qualification of (ii) as the adverse effects affecting the zones not involved in bilateral transactions are taken into account while solving MC problem. The actual flows do not disappear, however their transparency increases, and thus they become a part of the decision-making process of MC, which allocates the transfer capability to a set of competing market transactions. Considering the aforementioned remark we point out that the shape of zonal borders affects the market efficiency in the light of the existence of unscheduled flows.

There are many scientific approaches that can be used to identify loop flows but the methodology that suggest new

This work was supported by the EU and MSHE grant nr POIG.02.03.00-00-013/09. The use of the CIS computer cluster at NCBJ is gratefully acknowledged.

configuration of BZs based on loop flow analysis does not exist. This paper is aimed at fulfilling this gap and introduces a new method that addresses the problem. First, we identify a zone, the internal transactions of which cause the most of adverse effects in the system in terms of unscheduled flows. Second, we attribute each node of this zone with an amount of injected/withdrawn power which is responsible for isolated effect of unscheduled flow. This allows for spatial separation of the regions being average positive and negative contributors of the effect. Division line splits the zone into two BZs.

As the result, transactions previously categorized as internal, become visible as inter-zonal ones and start to be controlled by the Market Coupling. In order to achieve a new shape of the market, the method needs a certain bidding zone structure as a reference point. That is why the presented reasoning can be used on zonal structures resulting from the LMP or PTDF-based delimitations, as well as on structures that follow borders of countries or are defined by expert knowledge. The following sections present a mathematical background of the loop flow identification, clustering methods and an exemplary case study that illustrates the presented approach.

## II. METHODOLOGY

The whole process consists of several stages; market simulations allow to identify power flow on transmission lines in multiple scenarios of load patterns, this allows for estimating the magnitude of average loop flow effect. The zone responsible for the most of the inefficiency is chosen and clustered into two sub-regions in order to reduce the unscheduled power transmission. The following paragraphs introduce the details of the procedure.

### A. Load flow simulation

In order to achieve a set of nodal injections and withdrawals, market coupling algorithm [1] is utilized to determine accepted bids and offers, and, in consequence, nodal injections. To determine the power flows, we solve a DC power flow problem for these injections.

### B. Identification of loop flows

There are four possible categories of power transmission which seem to be important from the perspective of holding responsibility for utilizing the transmission infrastructure; (i) internal exchange, (ii) import/export, (iii) transit flow and (iv) loop flow. The basic understanding of the aforementioned terms in respect to attribution of nodes to different zones is the following:

i. internal exchange (IN) takes place if all, generator, load and transmission line are placed in the same zone,

ii. import/export (IE) refers to situation when generator and load are in different zones, but both ends of the line are attributed to at least one of these zones,

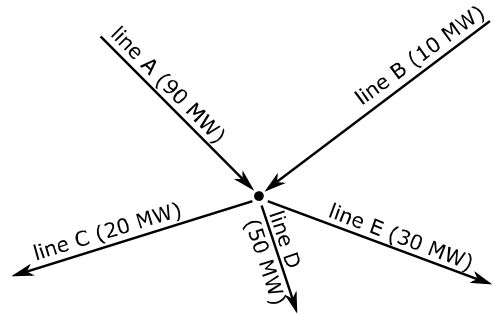

|        | to C | to D | to E |
|--------|------|------|------|
| from A | 18   | 45   | 27   |
| from B | 2    | 5    | 3    |

Figure 1. Assuming Proportional Sharing Principle leads to a conclusion that if in-flowing power on lines A and B equals 90 and 10 megawatts respectively, then 90% of power flowing through each of C, D & E comes from A and one tenth from branch B.

iii. transit (TR) occurs when line is operated by a zone which does not take part in the transaction and gen/load are in different zones,

iv. loop flow (LF) is the effect of power transfer through external infrastructure while both parties reside in the same region.

Obviously, in a real power system, a power flow in a given transmission line rarely can be categorized strictly into one of the four above classes, since the power flowing through the line usually cannot be attributed to one source (generation) node and one destination (load) node. Thus, we need to perform a decomposition of the power flowing in the line in order to divide it into components resulting from transfers between sets of loads and generators placed in the same zone, or from any pairs of different zonal attribution. Such an analysis demands determination of mutual interactions between sources and sinks participating in the power exchange.

Many authors devoted their studies to the subject of tracing the flow of electricity. This work is based on Bialek's Proportional Sharing Principle (PSP) [14], which along with lossless DC power flow analysis constitutes the main assumption of the study. The PSP can be summarized by the following statement: each node works as a "perfect mixer," which means that mutual proportion of in-flows is reflected by the components of out-flows (Fig. 1).

The method derived from the aforementioned rule has been improved by adding functionalities that allow for detailed power flow decomposition, i.e. exploring the magnitude of power flowing through each line as the result of exchange between each generator/load pair and categorizing flows' elements into the four classes defined above. Furthermore, the construction of all variables has been rephrased in the language of linear algebra, which allows for neat and more effective implementation.

For a given line $k$, one can find the matrix of mutual power exchange ($\mathbf{X}^k$) which means that node $i$ gets from generator $j$ power equal to $X^k_{ij}$ (cf. Appendix). Having this information, it is essential to ask which part of power transmission is actually a loop flow, and which constitutes more desirable forms of transaction-based power exchange. In order to categorize flows listed in matrices $\mathbf{X}$, one must discover interdependencies between zonal ascription of both ends of particular line and of the pair generator-load imposing transmission over the line.

The complete profile of the options (Fig. 2) allows to perform a decomposition in respect to each line of the system (below, line $k$ connects nodes $k_1$, $k_2$ and for each node $u$, operator $z(u)$ returns the zonal attribution of $u$).

$$X^k_{ij} \to \text{IN} \Leftrightarrow z(i) = z(j) = z(k_1) = z(k_2)$$

$$X^k_{ij} \to \text{IE} \Leftrightarrow z(i) \neq z(j) \wedge \{z(k_1), z(k_2)\} \in \{z(i), z(j)\}$$

$$X^k_{ij} \to \text{TR} \Leftrightarrow z(i) \neq z(j) \wedge \exists u \in \{i,j\}: z(u) \notin \{z(i), z(j)\}$$

$$X^k_{ij} \to \text{LF} \Leftrightarrow z(i) = z(j) \wedge \exists u \in \{i,j\}: z(u) \notin \{z(i), z(j)\}$$

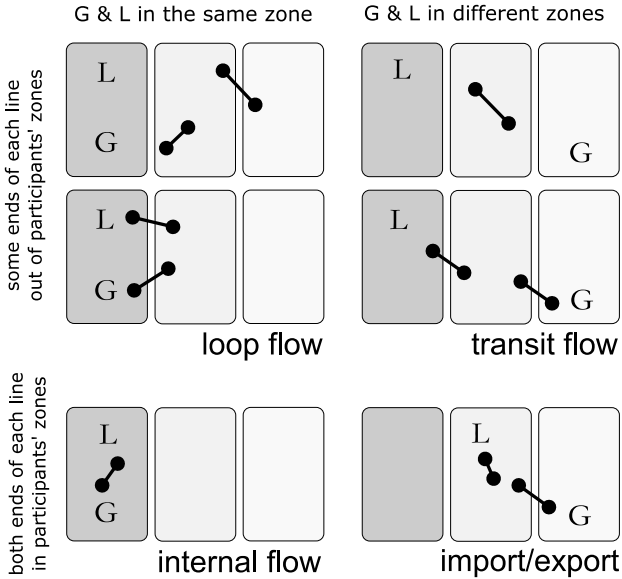

Figure 2. The category of power flow component is based on two indicators: zonal attribution of transaction parties (columns) and localization of transmission line (rows). The picture presents possible relations between the aforementioned. Different shades of figures indicate separate zones. Generation and load (G & L) refer to nodes of aforementioned source and sink denoted as $i, j$.

*C. Target zone – definition and clustering*

As the method aims at enhancing a zonal configuration by increasing number of zones, it starts with identification of one zone which can be treated as the greatest contributor to the loop flow phenomenon. The choice is made according to the ranking of the adverse effects of each zone's internal transactions. In this work, we decided to pay attention for the absolute values of flows, however relative approach (juxtaposing amount of flow in respect to the transmission capability of the line) is also an acceptable solution.

Once the influence of each node of the zone on the loop flow is determined by PF decomposition, hierarchic clustering [7] can be used to group nodes of the target zone. Our approach is to utilize values of power injections responsible for loop flows (and only them – cf. last equation of Appendix) as an input into the clustering routine. The area characterized by net positive "loop flow inducing" injections would be separated from region which behaves as net importer. As the result the zone responsible for the greatest amount of the unscheduled flow would be divided into two parts.

## III. CASE STUDY

This section brings the analysis of a simple test system (Fig. 3) used by Bialek in his introduction to PSP [14].

Let us assume that the zonal market identifies nodes 1 and 3 as one zone (A) whereas nodes 2 & 4 are located in other zone (B). Moreover, let the power flow described in [14] be the average power transmission over examined scenarios.

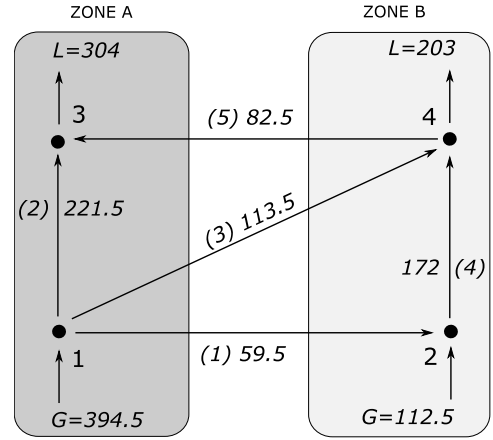

Figure 3. DC power flow solution for Bialek's exemplary test system. All values of power transfers and injections are given in megawatts. Each arrow indicates a direction of the exchange.

Decomposition of PF performed according to the described procedure gives the results listed in Table 1.

The implications based on data presented in the table are the following. "Target zone" is zone A, since most of (in this case all) the adverse effects turn out to be results of intra-zonal exchange between generator 1 and load 3 – which obviously is an intra-zonal transaction. This information can be obtained by the analysis of matrices $\mathbf{X}^k$ for each line $k$ involved in loop flow transmission. In this case, the only possible division of the target zone is to separate nodes 1 & 3, which does not leave much for a clustering analysis. The effect of decomposition made in the light of the improved zonal configuration can be found in Table 2.

As the result, all loop flows become recognized as transit flows. Moreover, previously internal exchange over the line 2 is now treated as IE across the border between zones B & C, established in the process of optimization.

| line | IN | IE | TR | LF |
|---|---|---|---|---|
| (1) | 0 | 42.31 | 0 | 17.19 |
| (2) | 221.50 | 0 | 0 | 0 |
| (3) | 0 | 80.70 | 0 | 32.80 |
| (4) | 79.99 | 74.82 | 0 | 17.19 |
| (5) | 0 | 32.51 | 0 | 49.99 |

Table 1. Initial decomposition of PF
(ZONE A: nodes 1 & 3, ZONE B: nodes 2 & 4)

| line | IN | IE | TR | LF |
|---|---|---|---|---|
| (1) | 0 | 42.31 | 17.19 | 0 |
| (2) | 0 | 221.50 | 0 | 0 |
| (3) | 0 | 80.70 | 32.80 | 0 |
| (4) | 79.99 | 74.82 | 17.19 | 0 |
| (5) | 0 | 32.51 | 49.99 | 0 |

Table 2. PF decomposition on optimized configuration
(ZONE A: node 1, ZONE B: nodes 2 & 4, ZONE C: node 3)

IV. CONCLUSIONS

The presented analysis is a consequence of the assumptions highlighted in sec. II. Although neglecting transmission losses may be treated as a reasonable simplification, the basic rule expressed by PSP ought to be discussed in the light of some concurrent premises. The main alternative which can be used for determining power flow decomposition is founded on the concept of superposition of power flows. The existence of counter-flows (two flows of opposite direction, which add up to one, visible "net" power flow) makes the model more complicated, however it allows for performing much deeper analysis of all possible interactions between selected generator and complete set of loads (or load against all generators). Particularly, a loop flow caused by two nodes: generator 2 and load 1 was not transparent according to the PSP, generator 2 introduces no contribution into line 2. The alternative assumption would result in increasing LF originating in zone 1 as well, however the fact that these two loops are of opposite direction, the superposition of effects could in fact be close to the one achieved by presented "net" effect. This does not change the fact that widely accepted Proportional Sharing Principle leads to satisfactory conclusions.

Results clearly indicate that by performing the optimization of bidding zone configuration one may expect a significant reduction of loop flow phenomenon. It is worth emphasizing that the physical flow does not need to change and yet the market works more efficiently. Eliminating unscheduled part of power transmission allows to reduce amount of power reserved for the unplanned transmission which, due to network security reasons, is often overestimated. Unblocked capacity serves for higher power exchange, further reduction of price differences and increase of social welfare on the whole market.

Future work should cover the comparison of different decomposition methods and testing variety of clustering techniques on large-scale power systems.

APPENDIX

Below we introduce an algebraic background of the described analysis. Let the system consist of $N$ nodes (buses) interconnected by $M$ lines. As the result of power flow analysis we get vector $\mathbf{f} = (f_1, ..., f_M)$ with coordinates equal to amounts of power exchanged through the systems' branches. Incidence matrix $\mathbf{G}$ ($N \times M$) consists of zeros except elements $G_{ki} = 1$ and $G_{kj} = -1$ if line $k$ is defined as a directed connection from node $i$ to node $j$. $\mathbf{G}$ can be decomposed into two $\mathbf{G_f}$, $\mathbf{G_t}$, useful in further definitions and consisting of non-negative elements only, so that

$$\mathbf{G} = \mathbf{G_f} - \mathbf{G_t}.$$

Next, we construct matrix $\mathbf{F}$, which is similar to directed adjacency matrix, but instead of zeros and ones it consists of amount of power flowing between the nodes, i.e. $F_{ij} = f_k$ means that $k^{\text{th}}$ line joins nodes $i$ and $j$ transmitting power $f_k$. If operator "diag" converts a vector into a square diagonal matrix so that $[\text{diag}(\mathbf{f})]_{ij} = f_i \delta_{ij}$, we can write

$$\mathbf{F} = \mathbf{G_f}^T \cdot \text{diag}(\mathbf{f}) \cdot \mathbf{G_t}.$$

A vector of total amount of power flowing through the nodes will be referred to as $\mathbf{p} = (p_1, ..., p_N)^T$, the elements of $\mathbf{p}$ can be defined as

$$p_i = \sum_{j=1}^{N} F_{ji} + p_i^g = \sum_{j=1}^{N} F_{ij} + p_i^l,$$

where $\mathbf{p}^g = (p_1^g, ..., p_N^g)$ and $\mathbf{p}^l = (p_1^l, ..., p_N^l)$ denote vectors of nodal generations and loads, both expressed in positive numbers.

Upstream and downstream approaches to the problem of PSP concern examining relations between amounts of power flowing in and out from the node, respectively. Figure 1 depicts a case, where in-flows are in proportion 9:1 (upstream, /u/) and out-flows – 2:5:3 (downstream, /d/). Matrices of flow contributions ($\mathbf{C}$) and flow distribution ($\mathbf{A}$) for both descriptions (u & d) introduced and interpreted by Bialek [14] may be formed as follows:

$$\mathbf{C_u} = \text{diag}^{-1}(\mathbf{p}) \cdot \mathbf{F},$$

$$\mathbf{A_u} = \mathbf{I} - \mathbf{C_u}^T,$$

$$\mathbf{C_d} = \mathbf{F} \cdot \text{diag}^{-1}(\mathbf{p}),$$

$$\mathbf{A_d} = \mathbf{I} - \mathbf{C_u}.$$

Distribution matrices can be used as operators converting nodal in/out flow into vectors of generation/load, however the greatest strength of this formalism is visible while calculating topological Generation Distribution Factors (GDFs) and Load Distribution Factors (LDFs):

$$\text{GDF} = \text{diag}\left(\Lambda(\mathbf{G_f C_u G_t}^T)\right)\mathbf{G_f A_u}^{-1},$$

$$\text{LDF} = \text{diag}\left(\Lambda(\mathbf{G_f C_d G_t}^T)\right)\mathbf{G_t A_d}^{-1},$$

where the operator $\Lambda$ takes a square matrix and outputs the a vector of its diagonal elements – $[\Lambda(\mathbf{F})]_i = F_{ii}$. The new distribution factors matrices allow to quantify the amount of power exchanged through a certain transmission line as the result of generator's injection (Generator-to-transmission line, G2T) or load's withdrawal (L2T):

$$\text{G2T} = \text{GDF}\,\text{diag}(\mathbf{p^g})$$

$$\text{L2T} = \text{LDF}\,\text{diag}(\mathbf{p^l})$$

Now it is possible to track all the pairs of loads and generators which influence each of system's transmission lines. For a given line $k$, one can find the matrix of mutual power exchange ($\mathbf{X}^k$) by

$$X^k_{ij} = \frac{1}{f_k} \text{L2B}_{ij} \cdot \text{G2B}_{ji},$$

which means that node $i$ gets from generator $j$ power $X^k_{ij}$.

Once the categorization of flows is finished, we get four vectors of different types of flows (cf. Table 1-2). Denoting loop flows by $\mathbf{f^{LF}}$, the nodal injections responsible for such flows ($\mathbf{p^{LF}}$) can be obtained by using the incidence matrix.

$$\mathbf{p^{LF}} = \mathbf{G}^T \mathbf{f^{LF}}.$$

This vector is a natural candidate for the set of values passed to the clustering algorithm (cf. sec II C). The target zone is identified by finding a zone with nodes causing the highest share of the effect.